\documentclass[10pt]{iopart}
\usepackage{graphicx}
\begin{document}
\title[Superradiant Light Scattering]{The Semiclassical and Quantum Regimes of Superradiant Light Scattering
from a Bose-Einstein Condensate}
\author{G.R.M. Robb\dag, N. Piovella\ddag\, and R. Bonifacio\ddag}
\address{\dag\ Department of Physics, University of Strathclyde, 107 Rottenrow, Glasgow G4 0NG, Scotland.}
\address{\ddag\ Dipartimento di Fisica, Universit\`a degli Studi di
Milano, INFM and INFN, Via Celoria 16, Milano I-20133, Italy.}
\date{last change by Nicola on {\tt \today}}
\eads{\mailto{g.r.m.robb@strath.ac.uk},
\mailto{nicola.piovella@mi.infn.it}}
\begin{abstract}
We show that many features of the recent experiments of Schneble
{\it et al.} [{\it D. Schneble, Y. Torii, M. Boyd, E.W. Streed,
D.E. Pritchard and W. Ketterle, Science {\bf 300}, 475 (2003)}],
which demonstrate two different regimes of light scattering by a
Bose-Einstein condensate, can be described using a one-dimensional
mean-field quantum CARL model, where optical amplification occurs
simultaneously with the production of a periodic density
modulation in the atomic medium. The two regimes of light
scattering observed in these experiments, originally described as
``Kapiza-Dirac scattering'' and ``Superradiant Rayleigh
scattering'', can be interpreted as the semiclassical and quantum
limits respectively of CARL lasing.
\end{abstract}


\section{Introduction}
The study of nonlinear optical phenomena arising from the
collective motion of atoms in dynamic optical fields has been an
active field of predominantly theoretical research over the last
decade. A large fraction of this work has been concerned with the
Collective Atomic Recoil Laser (CARL)
\cite{CARL:1,CARL:2,CARL:3,classical,Gatelli,CARL-MIT}. As part of
the continuing progress in the production and investigation of
ultracold atomic gases and Bose-Einstein condensates (BECs), there
have been several experiments which have demonstrated the validity
of these models and realized some of their predictions. Examples
include the observation of collective atomic recoil lasing by a
cold thermal gas in a high-finesse cavity
\cite{Zimmermann:1,Zimmermann:2} and the observation of
superradiant Rayleigh scattering by a BEC
\cite{Inouye,Kozuma,LENS}. Superradiant Rayleigh scattering
involves the production of pulses of coherently scattered
radiation simultaneous with the splitting of the condensate into
discrete momentum groups due to atomic recoil. Several theoretical
models have been used to describe the evolution of the
superradiant scattering process
\cite{classical,Gatelli,CARL-MIT,Meystre:1,Meystre:2,Ozgur,Trifonov},
including an extension of the original classical CARL model
\cite{CARL:1,CARL:2,CARL:3} to include a quantum treatment of the
atomic dynamics \cite{Gatelli,CARL-MIT,Meystre:2}. Recent
experimental work by Schneble {\it et al.} \cite{Schneble} has
shown that in addition to the superradiant Rayleigh scattering
process originally observed in \cite{Inouye}, there is a second
scattering regime termed ''Kapiza-Dirac scattering''. During
superradiant Rayleigh scattering, the scattering process involves
only emission of scattered photons i.e. absorption of pump photons
and emission of scattered (probe) photons. In contrast, during
Kapiza-Dirac scattering the scattering process involves both
emission {\em and absorption} of scattered (probe) photons i.e.
absorption of probe photons and emission of pump photons.

In this paper it is shown that many features of the recent
experiments of Schneble {\it et al.} \cite{Schneble} can be
described using a one-dimensional mean-field quantum CARL model,
where optical amplification occurs simultaneously with the
production of a periodic density modulation in the atomic medium.
Using this model, we demonstrate that the two regimes of
``Kapiza-Dirac scattering'' and ``Superradiant Rayleigh
scattering'' observed in \cite{Schneble},  can be interpreted as
the semiclassical and quantum limits respectively of CARL lasing.
It will be shown that the two regimes are distinguished by the
relative size of the gain of the superradiant scattering
process and the frequency separation of the absorption and
emission peaks. A significant difference between the results
presented here and those of other theoretical models of the
experiments of Schneble {\it et al.} \cite{Schneble,Meystre:3} is
that in this model the regime of scattering is not determined by
the pump pulse duration, so the scattering process does not evolve
in time from one to the other.

\section{Model}
\label{model} 
\begin{figure}[h]
\begin{center}
\includegraphics[width=6cm,clip=true]{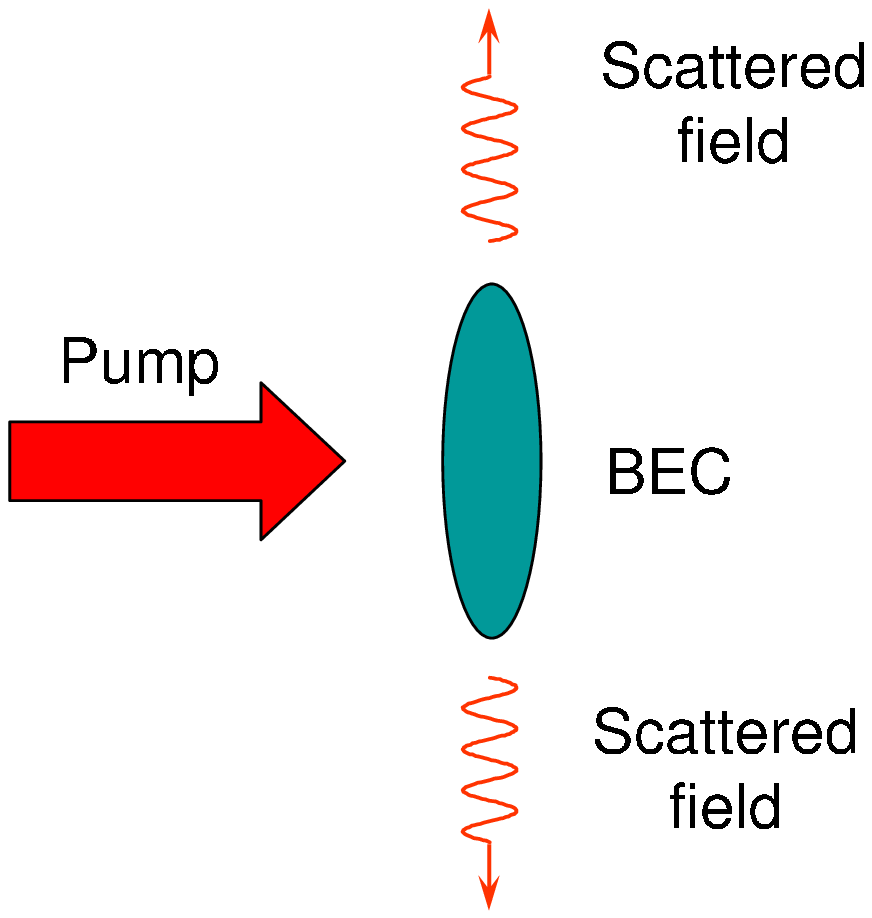}
\caption{Schematic diagram illustrating the geometry of the superradiant scattering
experiments of \cite{Schneble,Inouye}}
\end{center}
\label{fig1}
\end{figure}

The model used to describe the BEC-light interaction
is the mean-field quantum-CARL model originally derived in
\cite{Gatelli,CARL-MIT}. The model is one-dimensional and
describes the evolution of a backscattered (probe) field arising
from scattering of a pump laser field (assumed to be of constant
amplitude) by an elongated BEC. In the experiments of Schneble
{\it et. al.} \cite{Schneble}, the geometry of the experiment is
essentially two-dimensional, as illustrated schematically in
fig.~\ref{fig1} with emission of two endfire modes from each end
of the long axis of the condensate, propagating transversely to
the pump laser. If we assume that coupling between the endfire
modes (which are much weaker than the pump) is negligible, each
endfire mode can be assumed to evolve independently and the atomic
motion is one-dimensional.

When the pump laser is sufficiently detuned from the atomic
resonance, it leaves the atoms in the internal ground state.
Consequently radiation pressure due to absorption and subsequent
random incoherent, isotropic emission of a photon, can be
neglected. In this detuned regime, coherent scattering of the pump
laser is the dominant process. The atoms interact with a laser
beam of wave vector $\vec k$ and scatter photons of wave vector
$\vec k_s$, recoiling with a momentum $\hbar\vec q=\hbar (\vec k -
\vec k_s)$. The atoms, initially scattered randomly into various
momentum states, interfere with the atoms in the original momentum
state. This creates a matter wave grating having the correct
periodicity to further scatter the laser beam in the direction
$\vec k_s$. In an elongated condensate a preferential direction
for the scattered photons emerges causing superradiant Rayleigh
scattering. Both the matter wave grating and the scattered light
are coherently amplified \cite{Inouye,Schneble}.

In a simplified 1D description of the process along the direction
of the atomic recoil momentum $\hbar\vec q$, the evolution of the
matter wave field $\Psi(\theta,t)$ and of the dimensionless
amplitude $a(t)$ of the scattered radiation is determined by the
following quantum mean-field CARL model \cite{Gatelli,CARL-MIT}:
\begin{eqnarray}
i\frac{\partial\Psi}{\partial t}&=&-\omega_r
\frac{\partial^2\Psi}{\partial\theta^2}-ig \left[a
e^{i(\theta+\delta t)}- {\rm c.c.}\right]\Psi+\beta|\Psi|^2\Psi
\label{psi}\\
\frac{da}{dt}&=&gN \int d\theta|\Psi|^2e^{-i(\theta+\delta
t)}-\kappa a. \label{a}
\end{eqnarray}
where  $\theta=qz$ (with $q=|\vec q|\approx\sqrt{2}k$),
$a=(\epsilon_0 V/2\hbar\omega_s)^{1/2}E$ is the dimensionless
electric field amplitude of the scattered beam with frequency
$\omega_s$, $\omega_r=\hbar q^2/2m$ is the two-photon recoil
frequency, $g=(\Omega/2\Delta)(\omega d^2/2\hbar\epsilon_0
V)^{1/2}$ is the coupling constant, $\Omega=dE_p/\hbar$ is the
Rabi frequency of the laser field with constant amplitude $E_p$
and frequency $\omega=ck$, $\Delta=\omega-\omega_0$ is the
detuning from atomic resonance $\omega_0$,
$d=\hat\epsilon\cdot\vec d$ is the electric dipole moment of the
atom along the polarization direction $\vec\epsilon$ of the laser,
$V=AL$ is the volume of the condensate, $A$ is its cross-sectional
area, $L$ is its length, $N$ is the total number of atoms, and
$\delta=\omega-\omega_s$. The matter wave field is normalized such
that $\int_{-\infty}^{+\infty} d\theta |\Psi|^2=1$. The second
term on the right hand side of Eq. (\ref{psi}) is the
self-consistent optical lattice, resulting from the interference
between the laser and the scattered radiation, whose amplitude is
amplified by the matter wave grating described by the first term
on the right hand side of Eq. (\ref{a}). The third term on the
right hand side of Eq. (\ref{psi}) describes the mean-field effect
of the atom-atom interaction due to binary collisions, where
$\beta=8\pi\hbar q a_s N/mA$ and $a_s$ is the scattering length.
Eq. (\ref{a}) has been written in the 'mean-field' limit, which
models propagation of light with respect to the atoms by 
replacing the non uniform amplitude by its average value and by 
adding to the equation a damping term with a decay rate 
$\kappa\approx c/2L$ of the order of the inverse of the 
photon flight time through the condensate.

Notice that the interaction time $t$ appearing in Eqs. (\ref{psi})
and (\ref{a}) is the pump field duration, so that our model is
suitable to explore both Superradiant Rayleigh scattering ('long
pump pulse duration') and self-stimulated Kapitza-Dirac
diffraction regime ('short pump pulse duration') investigated in
ref. \cite{Schneble}. It will be shown here that the relevant parameter in these
experiments is not the pump pulse duration but the characteristic
time of the optical/matter-wave amplification process, given by 
the inverse of the gain rate.

If we assume that the atomic wave function is periodic on the
scale of the optical potential, with spatial period $2 \pi/q =
\lambda/\sqrt{2}$ which corresponds to a period in $\theta$ of $2
\pi$, then we can expand the atomic wave function in a Fourier
series
\begin{equation}
\label{expansion} \Psi(\theta,t) = \sum_{n=-\infty}^{\infty} c_n
(t) e^{i n (\theta+\delta t)}\label{Fourier}
\end{equation}
Furthermore, we assume that the condensate is sufficiently dilute
such that $4\pi\hbar a_s n_s/m\ll\omega_r$, where $n_s$ is the
average atomic density, so that the atom-atom interaction term in
Eq. (\ref{psi}) may be neglected. The effect of the atom-atom term
on the collective recoil lasing has been investigated in ref.
\cite{Amburgo}.

Substituting for $\Psi(\theta,t)$ using Eq.~(\ref{expansion}), it
can be shown \cite{Gatelli,CARL-MIT,Meystre:1} that Eqs.
(\ref{psi}) and (\ref{a}) can be rewritten as
\begin{eqnarray}
\frac{d c_n}{dt}  &=& - in(\omega_r n+\delta) c_n -
g\left( a c_{n-1} - a^* c_{n+1} \right) \label{cn1} \\
\frac{da}{dt} &=& gN\sum_{n=-\infty}^{\infty} c_n c_{n-1}^* -
\kappa a, \label{cn2}
\end{eqnarray}
In this quantum description, the Fourier expansion (\ref{Fourier})
is equivalent to expanding the wave function $\Psi(\theta,t)$ in
the set of momentum eigenstates with eigenvalues $\vec
p=(\hbar\vec q)n$ and $p_n=|c_n|^2$ is the probability for an atom
to have a momentum $\vec p=(\hbar\vec q)n$.

In the superradiant regime explored in the experiments of
ref.\cite{Inouye,Kozuma,LENS,Schneble}, the radiation damping rate
$\kappa$ is always much larger than the gain rate and/or the
recoil frequency $\omega_r$, so that the field amplitude follows
the atomic motion adiabatically . Hence, neglecting the time
derivative, Eq.(\ref{cn2}) yields
\begin{equation}
a\approx \frac{gN}{\kappa}\sum_{n=-\infty}^{\infty} c_n c_{n-1}^*.
\label{adia}
\end{equation}

\section{The semiclassical and quantum limits of the superradiant regime}
\label{linear} In order to obtain the gain coefficient of the
superradiant process in the semiclassical and quantum limits from
the dynamical equations, let us consider the initial equilibrium
state with no field, $a=0$, and all the atoms at rest, i.e. in the
momentum state $n=0$, with  $c_0=1$ and $c_m=0$ for all $m \neq
0$. Linearizing around this equilibrium solution and neglecting
the small detuning $\delta$ between the laser and scattered
frequencies, Eqs. (\ref{cn1}) and (\ref{adia}) reduce to the
single linear equation for $B=c_1+c_{-1}^*$:
\begin{equation}
\frac{d^2B}{dt^2}+\omega_r(\omega_r-iG)B=0, \label{B}
\end{equation}
where $G$ is the superradiant gain
\begin{equation}\label{G1}
    G=\frac{2g^2N}{\kappa}=\frac{\hbar\omega\Gamma\Gamma_{sc}N}{2AI_{sat}},
\end{equation}
$\Gamma$ is the natural linewidth,
$\Gamma_{sc}=\Gamma(\Omega/2\Delta)^2$ is the Rayleigh scattering
rate and $I_{sat}=c\epsilon\hbar^{2}\Gamma^2/4d^2$ is the
saturation intensity. Eq. (\ref{G1}) coincides with the
superradiant gain coefficient reported in Eq.(4) of
\cite{Schneble}.

It is easy to show that Eq.(\ref{B}) has an unstable solution
$B(t)\propto \exp[(\lambda_1+i\lambda_2)t]$, with
\begin{equation}
\lambda_1=
\left(\frac{\omega_r\sqrt{\omega_r^2+G^2}-\omega_r^2}{2}\right)^{1/2}
\label{gain}
\end{equation}
and $\lambda_2=\omega_rG/2\lambda_1$. Furthermore,
it is easy to show, from Eq.(\ref{B}), that
\begin{equation}
\frac{|c_{1}|}{|c_{-1}|}\approx
\frac{G/2}{\sqrt{(\lambda_1+G/2)^2+(\lambda_2+\omega_r)^2}}
\label{ratio}
\end{equation}
These expressions show that the instability may have a
semiclassical or quantum character, depending on the ratio between
the superradiant gain $G$ and the recoil frequency $\omega_r$. In
fact, in the limit where $G\ll\omega_r$, Eqs. (\ref{gain}) and
(\ref{ratio}) give $\lambda_1\approx G/2$,
$\lambda_2\approx\omega_r$ and $|c_1|/|c_{-1}|\approx
G/(4 \omega_r) \ll 1$. In this limit, the SR process is quantum in
nature, with only the lower state $n=-1$ being populated, and the
SR gain is $G$ \cite{Gatelli,Meystre:1}. In the opposite limit in
which $G\gg\omega_r$, Eqs. (\ref{gain}) and (\ref{ratio}) give
$\lambda_1\approx\lambda_2\approx\sqrt{\omega_rG/2}$ and
$|c_1|/|c_{-1}|\approx 1-\sqrt{2\omega_r/G}$. In this limit, the
SR process is semiclassical in nature, with the states $n=1$ and
$n=-1$ almost equally populated. In this case, the superradiant
gain is given by \cite{classical}
\begin{equation}\label{G2}
    G'=\sqrt{2\omega_rG}=2g\sqrt{\frac{\omega_rN}{\kappa}}.
\end{equation}
The SR gain in the semiclassical limit is always lower than the SR
gain in the quantum limit, since $G'/G=\sqrt{2\omega_r/G}\ll 1$.
Notice also the different dependence of the SR gain on $N$ in the
quantum and the classical limits.

\section{Numerical Results}
\label{numerics} In order to observe the behavior of the
collective scattering process in the nonlinear regime, Eqs.
(\ref{cn1}) and (\ref{cn2}) were integrated numerically with
initial conditions $a=0$, $c_{-1}=1/\sqrt{N}$, $c_0=\sqrt{1-1/N}$,
and $c_{m}=0$ when $m \neq -1,0$ and with parameters corresponding
to the semiclassical and quantum regimes of evolution. These
parameters correspond to the experiments of \cite{Schneble} i.e. a
$^{87}$Rb condensate illuminated by a pump beam of wavelength
$\lambda=780\mbox{nm}$ and an intensity of $63 \mbox{mW
cm$^{-2}$}$. The pump couples to the $5S_{1/2} \rightarrow
5P_{3/2}$ transition which has a natural width $\Gamma=0.37\times
10^8 $ s$^{-1}$, dipole moment $d=2.07\times 10^{-29}$ C$\cdot$m,
saturation intensity $I_{sat}=2.5 \mbox{mW cm$^{-2}$}$ and recoil
frequency $\omega_r=4.7 \times 10^4$ s$^{-1}$. The condensate had
a cigar-shaped form, $15\ \mbox{$\mu$m}$ in diameter and $200
\mbox{$\mu$m}$ in length, so that $\kappa=7.5\times 10^{11}$
s$^{-1}$. Using these parameters, the quantum and classical
superradiant gain coefficients of Eqs. (\ref{G1}) and (\ref{G2})
are $G\approx 4.9\times 10^6\times N/ |\Delta|^{2}$ and $G'\approx
6.8\times 10^5\times \sqrt{N}/ |\Delta|$, respectively, where
$\Delta$ is the pump-atom detuning in MHz. We assume that
$\delta=0$ and that half of the atoms in the condensate
participate in each of two superradiant emissions along the main
axis of the condensate and that the number of atoms participating
in the collective scattering process is $N=10^{5}$ rather than
$N=10^6$ as quoted for the number of atoms in the condensate in
\cite{Schneble}. Qualitative support for this assumption is
provided by fig.3A of \cite{Schneble}, which shows a large
fraction of the condensate atoms do not participate in the
coherent superradiant scattering process. From \cite{Schneble},
the 'Kapiza-Dirac' experiment was carried out using $\Delta = -420
\mbox{MHz}$, so that $g=3.2\times 10^6$ s$^{-1}$ and the quantum
superradiant gain is $G=3\times 10^6$ s$^{-1}$. Consequently the
collective scattering process is semiclassical in nature, since
$G/\omega_r\sim 58$. For the experiments in the 'superradiant
Rayleigh scattering' regime, $\Delta = -4400 \mbox{MHz}$,
$g=3.07\times 10^5$ s$^{-1}$ and the quantum superradiant gain
coefficient is $G=2.5\times 10^4$ s$^{-1}$, so that the collective
scattering process is quantum-mechanical in nature, since
$G/\omega_r\sim 0.53$.

\begin{figure}[h]
\begin{center}
\includegraphics[width=6cm]{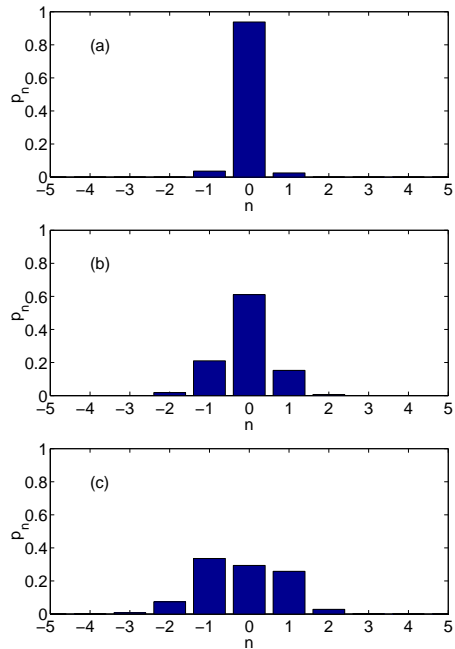}
\caption{Atomic momentum distribution in the semiclassical regime
with  $G=\, 58\omega_r$, when (a)  $t=11 \mu$s, (b) $t=15 \mu$s
and (c) $t=17 \mu$s.
 }
\label{kapiza_snaps}
\end{center}
\end{figure}

\begin{figure}[h]
\begin{center}
\includegraphics[width=8cm]{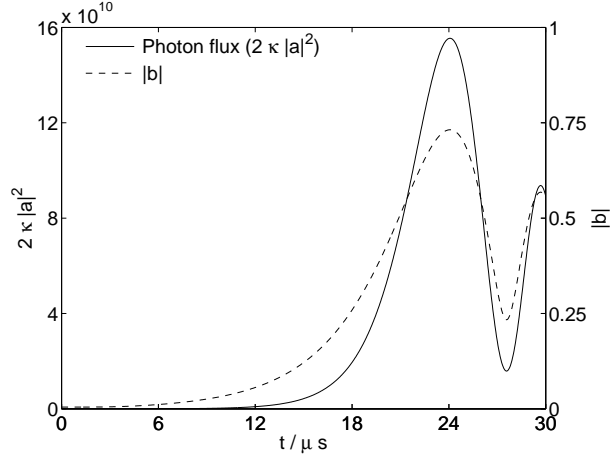}
\caption{Flux of scattered photons, $2\kappa|a|^2$, and bunching
factor, $|b|$, as a function of time in the semiclassical regime
with  $G=\, 58\omega_r$.} \label{fielda}
\end{center}
\end{figure}

Fig.~\ref{kapiza_snaps} shows snapshots of the atomic momentum
distribution at different times for the semiclassical case with
$\Delta = -420 \mbox{MHz}$. It can be seen from
fig.~\ref{kapiza_snaps}(a) that, in agreement with our theory, at
the beginning of the interaction momentum states $n=1$ {\em and}
$n=-1$ are the only non-zero momentum states to have significant
population, with $p_1 \sim p_{-1}$, where $p_n = |c_n|^2$. Eq.~(\ref{ratio})
predicts a ratio $|p_{1}|/|p_{-1}| \approx 0.7$. As time
progresses, many momentum states with both positive {\em and}
negative $n$ are populated. However, the momentum distribution is
not symmetric about $n=0$. The average momentum is less than zero,
so from momentum conservation there is a net gain of the scattered
radiation field, shown in fig~\ref{fielda}, due to the difference
in photon absorption and emission rates. Fig.~\ref{fielda} shows
that amplification of the scattered field in the semiclassical
case is simultaneous with the growth of a strong density
modulation in the condensate, as represented by the bunching
parameter, $b\equiv\langle e^{-i\theta}\rangle=\sum_n
c_nc_{n-1}^*$. The atomic momentum distributions shown in
fig.~\ref{kapiza_snaps} are similar to the time of-flight images
observed for the so-called ``Kapiza-Dirac scattering'' observed in
\cite{Schneble}, where population of momentum states due to
absorption {\em and} emission of radiation was observed. 


\begin{figure}[h]
\begin{center}
\includegraphics[width=6cm]{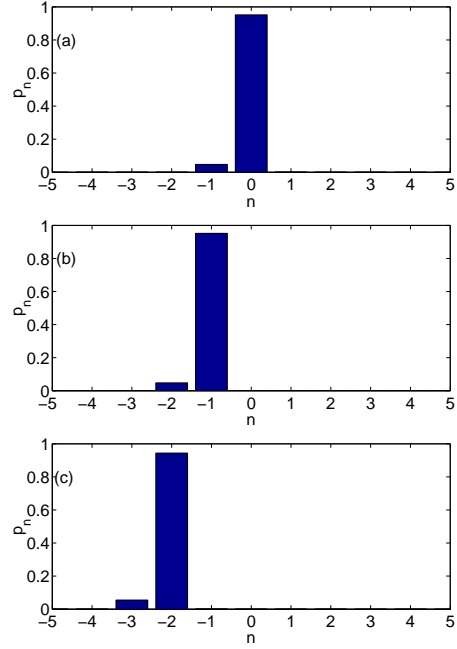}
\caption{Atomic momentum distribution in the quantum regime with
$G=0.53\omega_r$ when (a) $t=0.3$ms, (b) $t=1.0$ms and (c)
$t=1.7$ms.} \label{srs_snaps}
\end{center}
\end{figure}

\begin{figure}[h]
\begin{center}
\includegraphics[width=8cm]{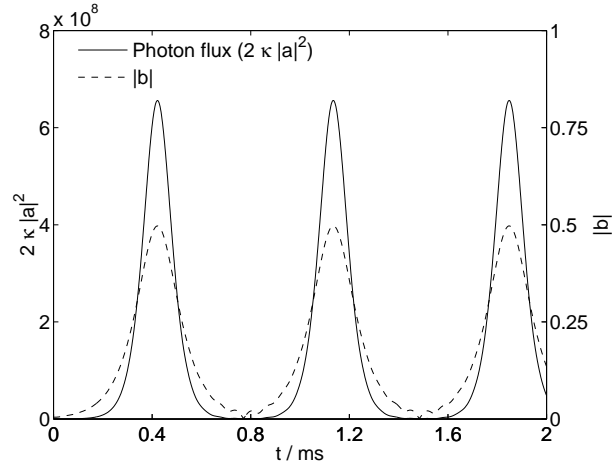}
\caption{Flux of scattered photons, $2\kappa|a|^2$, and bunching
factor, $|b|$, as a function of scaled time in the quantum regime
with $G=0.53\omega_r$.} \label{fieldb}
\end{center}
\end{figure}

Fig.~\ref{srs_snaps} shows snapshots of the atomic momentum
distribution at different times for the quantum case with $\Delta
= -4400 \mbox{MHz}$. It can be seen from fig.~\ref{srs_snaps}(a)
that, in agreement with our model, at the beginning of the
interaction momentum state $n=-1$ is the only non-zero momentum
state to have significant population, and that the population $p_1
\approx 0$. Eq.~(\ref{ratio}) predicts a ratio $|p_{1}|/|p_{-1}| \approx 0.01$.
As time progresses, the atomic population moves
sequentially from $n=-1 \rightarrow n=-2 \rightarrow ...$ and
states with $n>0$ are never populated. This sequential decrease in
atomic momentum gives rise to amplification of the scattered
radiation field and again occurs simultaneously with the
development of a strong density modulation in the condensate, as
shown in fig.~\ref{fieldb}. In contrast to the semiclassical case,
however, this amplification of the scattered field occurs due to
emission of scattered (probe) photons {\em only}, and the spread
in atomic momenta is much smaller than in the semiclassical case.
The atomic momentum distributions shown in fig.~\ref{srs_snaps}
are similar to the time of-flight images observed for
``Superradiant Rayleigh scattering'' in \cite{Schneble,Inouye},
where the atoms attain momentum {\em only} in the direction of the
pump beam  in discrete units of $\hbar \vec q$.

\section{Interpretation}
\label{interpretation} In \cite{Schneble}, the
Kapiza-Dirac scattering regime and the Superradiant Rayleigh
scattering regime are described as the ``short-pump pulse'' and
``long-pump pulse'' limits respectively. A problem with this
classification is that it implies that for sufficiently long pump
pulses the collective scattering makes a transition from
Kapiza-Dirac scattering to superradiant Rayleigh scattering, or in
the terminology of this paper a transition from the semiclassical
to the quantum regime of CARL lasing. The results presented here
however do not support this interpretation. They suggest that the
distinguishing feature between the two experiments in
\cite{Schneble} is not the pump pulse duration, but the atom-field
detuning (which differs by an order of magnitude in the
experiments of \cite{Schneble}), which determines the atom-field
coupling and consequently the timescale of the superradiant
scattering process. Rather than the ratio of the pump pulse
duration relative to the two-photon recoil time (the inverse of
the two-photon recoil frequency),
it is the size of the superradiant decay time relative to the
two-photon recoil time which determines whether or not the
scattering consists of a sequence of emission processes, as
observed in the ``superradiant Rayleigh scattering'' regime
\cite{Schneble,Inouye}, or simultaneous emission and absorption
processes, as observed in the ``Kapiza-Dirac regime''.

A simple justification for this argument is as follows: The atoms
are assumed to all be in the condensate initially, so that
$p_0=1$. The frequencies at which the probability of absorption
($n=0 \rightarrow n=1$) and emission ($n=0 \rightarrow n=-1$) are
maximum are non-degenerate and differ by a frequency of $2
\omega_r$.
A transition from ($n=0 \rightarrow n=-1$) will initiate
superradiant or superfluorescent decay. The characteristic time of
the superradiant decay  is $\tau_{sr}\sim G^{-1}$ for $\kappa\gg
g\sqrt{N}$ \cite{CARL-MIT}, so the spectral width of the
superradiant pulse is $\sim G=2g^2N/\kappa$. If the spectral width
of the SR pulse is much less than the absorption-emission
frequency shift i.e. $g^2 N/\kappa\le\omega_r$, then absorptive
transitions will not occur because they are non-resonant and only
($n=0 \rightarrow n=-1$) transitions, i.e. absorption of pump
photons and emission of probe photons, will occur. Consequently
the system evolves in the quantum CARL limit when $G\le\omega_r$.
In contrast, if the spectral width of the SR pulse is sufficiently
large (i.e. SR decay is sufficiently rapid) that it is much larger
than the absorption-emission frequency difference, so that $G\gg
\omega_r$, then absorptive transitions ($n=0 \rightarrow n=1$)
will be resonant and both emission and re-absorption of probe
photons will occur. Consequently the system evolves in the
semiclassical CARL limit when $G\gg\omega_r$. It should be noted
that in contrast to previous explanations
\cite{Schneble,Meystre:3} of the experimental results in
\cite{Schneble}, in our argument the duration of the pump pulse is
not a significant factor. The reason that the semiclassical CARL
or ``Kapiza-Dirac'' regime can be observed using a short pump
pulse in \cite{Schneble} is because the timescale of semiclassical
superradiant decay, $\tau'_{sr}\sim 1/G'$ (see  Eq.(\ref{G2})) is
much shorter than in the ``Superradiant Rayleigh Scattering''
example as a result of the decreased pump-atom detuning. In
Ref.~\cite{Schneble} a suppression of quantum SR gain  $G$ of
Eq.~(\ref{G1}) by around two orders of magnitude was observed and
attributed in \cite{Schneble} to the short duration of the pump
pulse. The CARL model described here explains the reduced gain
observed in \cite{Schneble} as a result of the fact that the SR
scattering process is evolving semiclassically. Consequently the
SR gain is not given by $G$ as given by Eq.~(\ref{G1}) but the
semiclassical SR gain $G'$ as given by Eq.~(\ref{G2}). The
semiclassical gain, $G'$, gives a value consistently smaller than
the one predicted by the quantum superradiant gain, $G$. Using the
same parameters as those used in figs~\ref{srs_snaps} and
\ref{fielda}, the ratio is $G'/G \approx 0.16$.


\section{Conclusion}
It has been shown that many features of the recent experiments of
Schneble {\it et al.} \cite{Schneble}, which show two different
regimes of light scattering by the BEC, can be described using a
one-dimensional mean-field quantum CARL model. The two regimes of
light scattering observed in \cite{Schneble}, described as
``Kapiza-Dirac scattering'' and ``Superradiant Rayleigh
scattering'' in \cite{Schneble}, can be interpreted as the
semiclassical and quantum limits respectively of CARL lasing. In
the semiclassical limit, when $g^2N/\kappa\gg\omega_r$, the
collective scattering process involves both absorption and
emission of probe photons, and for sufficiently long times many
momentum states are populated simultaneously. In the quantum
limit, however, when $g^2N/\kappa\le\omega_r$, the collective
scattering process involves emission of probe photons only, and a
maximum of two momentum states are populated at any time. We
provide a simple explanation of these results in terms of a
comparison between the frequency separation of probe emission and
absorption events and the spectral width of superradiant decay
between the initial and recoiling condensates.
In contrast to previous models of the
experiments in \cite{Schneble}, the pump pulse duration is not a
significant factor in our interpretation. The results presented
here support the view of the BEC-light interaction as a strongly
coupled atom-optical system where the atoms and light (and
consequently the matter-wave and optical gratings) evolve
dynamically and self-consistently. This picture gives a more
correct and complete description of the interaction than models in
which the dynamics of either the atoms (matter-wave grating) or
the optical fields (optical grating) are neglected.

\end{document}